\documentstyle[aps,prl]{revtex}
\begin{document}
\title{ Reply to ``Comment on 
`Inverse exciton series in the optical decay of 
an excitonic molecule'\ " }
\draft
\author{Eiji Tokunaga,$^1$ A.L. Ivanov,$^{2}$ 
Selvakumar V. Nair,$^1$ and Yasuaki Masumoto$^{1,3}$}
\address{$^1$ERATO Single Quantum Dot Project, Japan Science and 
Technology Corporation, 5-9-9 Tokodai, Tsukuba 300-2635, Japan}
\address{$^2$ TCM, Cavendish Laboratory, University of Cambridge, UK}
\address{$^3$Institute of Physics, University of Tsukuba, 
Tsukuba 305-8571, Japan}
\date{\today}
\maketitle

\begin{abstract}
As a reply to the Comment by I.S. Gorban {\it et al.} (Phys. Rev. B, 
preceding paper) we summarize our criticism on their claim of 
the first observation of the $M$ series in $\beta$-ZnP$_2$.  
We support our analysis by reporting the first observation of inverse 
{\it polariton} series from the excitonic molecules selectively 
generated at ${\bf K}_m \simeq {\bf 0}$ in a CuCl single crystal. 
This observation and its explanation within the bipolariton model 
complete our proof of the biexcitonic origin of the inverse series. 
\end{abstract}


\narrowtext
\twocolumn

A monocrystal CuCl is a prototype material in the physics of 
excitonic molecules, due to the polarization isotropy of the optical 
transitions to the excitonic states and due to the large binding energy 
of the molecules $\epsilon^m \simeq$ 32~meV \cite{Ueta}. In a recent 
paper \cite{Tokunaga} we reported the observation of the $M$ series, which 
originates from the optical decay of the molecules into the excitonic 
states $1s$, $2s$, $3s$, $4s$, and $\infty s$. The lines $M_{n \geq 
2s}$ are extremely weak and their detection requires a very modern 
spectroscopic technique and a high-quality CuCl single crystal. 
In our paper we explained briefly why the claim by Gorban {\it et al.} 
\cite{Gorban2,Gorban3,Gorban4} to the observation of the $M$ series in 
$\beta$-ZnP$_2$ is not justified (see reference [17] of 
Ref.~\cite{Tokunaga}). By their Comment \cite{Gorban1} the authors force 
us to present a more detailed criticism of their experiments. 

Monoclinic zinc diphosphide ($\beta$-ZnP$_2$) is an optically 
highly-anisotropic semiconductor with eight molecules per unit cell, 
which give rise to $72$ phonon modes. As a result, the excitonic 
spectra are very complicated and strongly depend on the excitation 
and polarization geometry. Thus the correct interpretation of the 
optical spectra of $\beta$-ZnP$_2$ requires a very accurate experiment, 
a high-quality sample, and a critical analysis of the experimental 
data. The first reports \cite{Syrbu} on the observation of the inverse 
hydrogen-like series in $\beta$-ZnP$_2$ attribute the series to a 
bielectron-impurity complex. To the best of our knowledge, these 
results have never been confirmed. The next reports 
\cite{Gorban2,Gorban3} on the inverse series ascribe the 
photoluminescence (PL) spectra of highly-excited $\beta$-ZnP$_2$ to 
the $M$ series. Even without commenting on Gorban {\it et al.}'s 
experiments, 
we will show below that their interpretation 
of the PL spectra in terms of the $M$ series 
is self-contradictory and incorrect. 

Firstly, from their experimental data Gorban {\it et al.} conclude that 
the biexciton binding energy $\epsilon^m$ in $\beta$-ZnP$_2$ is 
unusually high, i.e., $\epsilon^m = 14.9$~meV ($0.32$ of the 
excitonic Rydberg $\epsilon^x$). In order to explain this result the 
authors claim without any justifications that the relevant electron and 
hole effective masses in $\beta$-ZnP$_2$ are given by $m_e = 1.7 m_0$ and 
$m_h = 21.3 m_0$ ($\sigma = m_e/m_h = 0.08$) \cite{Gorban3}. This 
strongly contradicts the Kane model, which basically requires $m_h 
\sim m_0$, and the first experiments \cite{Yu} on excitonic polaritons 
in $\beta$-ZnP$_2$ which yield the effective excitonic mass 
$M_x = m_e + m_h \leq 3 m_0$. The latest high-precision experiments 
by two-photon and magneto-optical spectroscopies \cite{Froehlich} 
indicate the high anisotropy of $\beta$-ZnP$_2$ single crystals 
with $m_e = 0.7 m_0$, $m_h = 1.1 m_0$ along the $a$-axis, $m_e = 1.1 m_0$, 
$m_h = 1.4 m_0$ along the $b$-axis, and $m_e = 0.18 m_0$, $m_h = 0.20 m_0$ 
along the $c$-axis, respectively. These values clearly demonstrate that 
the effective masses chosen by Gorban {\it et al.} are 
irrelevant. Furthermore, with reference to $\sigma = 0.08$  
the authors claim that their $\epsilon^m = 0.32 \epsilon^x$ 
``corresponds to the variational calculations of 
Refs.~\cite{Akimoto,Rice}'' (see p.~339 of Ref.~\cite{Gorban3}). 
This is a misinterpretation of the theoretical results of 
Refs.~\cite{Akimoto,Rice}. An inspection of Fig.~1 of 
Ref.~\cite{Akimoto} and Fig.~3 of Ref.~\cite{Rice} shows that 
the mass ratio $\sigma = 0.08$ yields $\epsilon^m \simeq 0.12 
\epsilon^x$. As a result, even for the above incorrect value of the ratio 
$m_e/m_h$ the molecule binding energy should be only $\epsilon^m = $
5.64~meV rather than 14.9~meV.  

Secondly, the intensity ratio between two main replicas of the series 
reported by Gorban {\it et al.} is too high to be ascribed to the $M$ 
series. The intensity ratio of two lines marked by $M_1$ 
and $M_2$ in Fig.~1 of Ref.~\cite{Gorban3} is given by $I_{M_2}/I_{M_1} 
\simeq 1/3$ ($I_{M_1}=I_{M_{1T}}+I_{M_{1L}}$). 
In Ref.~\cite{Gorban4} the authors explain that the above 
ratio should be corrected due to the saturation of the $M_{1T}$ line 
and that the true value is $I_{M_2}/I_{M_{1}} \simeq 1/9.08$ as shown in 
the intensity dependence in Fig.~2 of Ref.~\cite{Gorban3}. The 
bipolariton model \cite{Ivanov} with the Akimoto-Hanamura trial 
wave function \cite{Akimoto} yields $I_{M_2}/I_{M_1} \simeq 1/5000$ for 
excitonic molecules in CuCl (the electron-hole mass ratio $\sigma = 
0.28$). This value is in agreement with our experiments, which indicate  
$I_{M_2}/I_{M_1} \simeq 1/7000$. 
Using a precise wave function of the hydrogen 
molecule \cite{Kolos}, we find $I_{M_2}/I_{M_1} \simeq 1/800$ in 
the limit $\sigma=0$. Thus, within the uncertainty in the value of 
$\sigma$ in $\beta$-ZnP$_2$, we estimate, for excitonic molecules in this 
semiconductor, $I_{M_2}/I_{M_1} \leq 1/1000$. The latter value is at 
least two orders of magnitude smaller than that reported by Gorban 
{\it et al.} \cite{Gorban2,Gorban3}. Note that in the Comment 
\cite{Gorban1} the authors have corrected their intensity 
ratio as $I_{M_2}/I_{M_1} \simeq 1/230$ referring to 
the $new$ intensity dependence in Fig.~2 of the Comment, but 
this intensity dependence contradicts that reported 
in Fig.~2 of Ref.~\cite{Gorban3}. 

The low-temperature PL spectra of $\beta$-ZnP$_2$ single crystals 
at moderate optical excitations have been studied 
in Ref.~\cite{Kondo}. According to Ref.~\cite{Kondo}, 
the $M_1$ emission band indeed develops at an optical excitation power of 
about 10~kW/cm$^2$, however, without any indications of the $M$ series. 
Instead, below the $M_1$ band another PL line arises (at 1.523~eV), 
which grows quadratically with the excitation intensity and has a 
comparable intensity with the $M_1$ band. This $M$-like emission line 
is attributed to excitonic molecules trapped by surface defects, 
because a similar spectral line is presented in the reflection spectra 
but is absent in the absorption spectra. Furthermore, the line is not 
reproducible from sample to sample but found only for a few samples. 
Note that the above described 
$M$-like emission at 1.523~eV exactly coincides with the PL line 
labeled by $M_2$ in Fig.~1 of the Comment (see also Fig.~1 of 
Ref.~\cite{Gorban3}). 
According to Ref.~\cite{Goto}, the molecule energy is 
3.1102 $\pm$ 0.0002~eV. Therefore the inverse $M_2$ emission 
should arise at 1.519~eV, rather than at 1.523~eV 
in the Gorban {\it et al.}'s data. 

Because of its extremely weak intensity, 
the confirmation of the inverse series requires 
many severe experimental checks. 
The two-photon resonant generation of excitonic molecules with 
translational momentum $\hbar {\bf K}_m$ is crucial for unambiguous 
discrimination of the $M$ series in our experiments. 
In Fig.~1 we compare the PL spectra of 
a high-quality CuCl single crystal (sample $No.~1$) 
recorded under the selective excitation of the molecules at 
${\bf K}_m = 2 {\bf k}_0$ (solid curve) and 
under the interband excitation (dashed curve). 
Here ${\bf k}_0$ is the wave vector of the 
pump polaritons. In the latter case 
the $M_T$ and $M_L$ emission lines are absent 
because of weak cw-laser excitation, but only $1s$ exciton emissions and 
the relavant bound exciton lines are present. 
As shown by the solid curve, 
even under the two-photon excitation of the molecules, 
bound exciton structures similar to those in the dashed curve 
grow quadratically with the excitation intensity 
together with the $M$ series. 
They originate from the secondary excitons 
generated by the molecule decay. 
As shown by the dashed curve,  
an emission from impurity-bound excitons occurs at 3.082~eV. 
Therefore, the line marked by $M_{Z_{1,2}}$ cannot be attributed to 
the inverse series, 
although its energy coincides with the molecule emission 
leaving the longitudinal $Z_{1,2}$ exciton behind \cite{Hasuo}. 
On the other hand, 
our samples are free from any impurity emissions at the 
spectral position of the $M_{2s}-M_{\infty s}$ lines. 
Thus we assign these lines to the inverse $M$ series. 
Furthermore, we checked the reproducibility of the $M_{2s,3s,4s}$ series 
for several high-quality samples of a CuCl single crystal. Figure 2 shows the 
PL spectrum of sample $No.~2$ under resonant generation of excitonic 
molecules at ${\bf K}_m = 2 {\bf k}_0$. The period of interference 
pattern of sample $No.~2$ is different from that of sample $No.~1$. 
However, the characteristics of the $M$ series are identical for 
both samples. 

At the present time we have improved the experimental setup and can observe 
the optical decay of excitonic molecules selectively excited at 
various ${\bf K}_m$ ($0 \leq K_m \leq 2 k_0$). Figure 3 shows the 
inverse emission series from excitonic molecules with ${\bf K}_m \simeq 
{\bf 0}$. Now the molecule cannot decay into the pure excitonic states 
$1s$, $2s$, ..., but resonantly dissociates into two outgoing polaritons 
associated with the corresponding optical transitions. Thus the $M$ 
series of Fig.~3 is described as an inverse {\it polariton} series. 
Because the main contribution to the series is due to 
the resonant dissociation of the molecule 
into the lower dispersion branch polaritons, 
we designate the series by $LP_{ns}$. Note that the intensity ratio 
$I_{LP_{2s}}/I_{LP_{1s}}$ is about $1/100$, i.e., by two orders of 
magnitude larger than $I_{M_{2s}}/I_{M_{1s}}$ corresponding to the 
molecules initially photogenerated at ${\bf K}_m = 2{\bf k}_0$ 
\cite{Tokunaga}. With the bipolariton model of an excitonic molecule 
\cite{Ivanov}, both of the inverse series, $M_{ns}$ of Fig.~1 and $LP_{ns}$ 
of Fig.~3, are explained self-consistently using the same trial molecule 
wave function \cite{Tokunaga2}. This completes our proof of the 
biexcitonic origin of the inverse series observed in CuCl 
single crystals. The use of a weak selective resonant excitation of the 
molecules allows us to avoid the nonlinear reabsorption processes and 
a thermal distribution of the molecules which considerably changes 
the intensity ratios of the inverse series. 

In their experiments\cite{Gorban2,Gorban3,Gorban1}, Gorban {\it et al.} 
have used only the interband excitation of $\beta$-ZnP$_2$. Such an 
excitation is not suitable for the discrimination of the $M$ series 
as we have shown above. Furthermore, while the authors do not observe 
their $M$ series at moderate interband excitations of $\beta$-ZnP$_2$ 
(Ar-laser of a few kW/cm$^2$ intensity), they describe a 
corresponding rich PL spectrum mainly in terms of the optical 
decay of nonlocalized excitonic molecules (see the lines marked 
by $M_1$, $C$, and $G$ in Fig.~3 of Ref.~\cite{Gorban3}). 
According to Gorban {\it et al.}, the $G$-band is due to the 
degenerate two-photon radiative annihilation of biexcitons 
with ${\bf K}_m \simeq {\bf 0}$ and the $C$-line is due to a 
``dielectric liquid'' of excitonic molecules. We consider this 
interpretation as absolutely speculative and unjustified. 
In particular, the $M_{1s}$ band of thermally distributed excitonic 
molecules is not expected to show any fine structure due to 
the degenerate two-photon annihilation of the molecules \cite{Ueta,Anzai}. 
Finally, if the nonlocalized biexcitons are mainly responsible 
for the complicated PL spectrum plotted in Fig.~3 of Ref.~\cite{Gorban3}, 
as the authors incorrectly interpret, it is not understandable 
why in this case they do not observe their $M$ series. 

Once the inverse $M$-series is observed \cite{Tokunaga}, the 
question on the reconstruction of the internal molecule wave function 
naturally arises. Here, the relevant theoretical questions are how the 
molecule decays radiatively to the $ns$ exciton states and how the relative 
intensities of the $M_{ns}$ replicas relate to the molecule wave function. 
The theory of the optical decay of an excitonic molecule presented 
by Gorban {\it et al.} \cite{Gorban4} is incorrect. 

Firstly, the use of Wang's trial wave function, Eq.~(1) of 
Ref.~\cite{Gorban4}, means that the biexciton is identified with the 
hydrogen molecule. The Wang wave function is suitable for H$_2$ due 
to the adiabatic approximation, which requires $(m_e/m_h)^{1/4} \ll 1$. 
The latter inequality definitely does not hold even for the incorrect 
ratio $m_e/m_h = 0.08$ used by the authors for biexcitons in $\beta$-ZnP$_2$. 
In other words, an envelope wave function for the relative motion of 
two constituent excitons is absent in Eq.~(1) of Ref.~\cite{Gorban4}. 
However, the authors ignore the above standard arguments and 
take the results of the well-known calculations \cite{Wang} on the 
hydrogen molecule with the same values of the fitting parameter 
($Z=1.166$) and binding energy ($0.31$ Rydberg). The Wang wave function 
is irrelevant for excitonic molecules. The corresponding evaluation of 
the molecule binding energy within the four-particle Schr\"odinger 
equation, claimed by the authors, cannot be reproduced. 

Secondly, Eq.~(2) of Ref.~\cite{Gorban4} proposed for the intensities 
of the $M_{ns}$ replicas is incorrect, because (i) two of the arguments 
of the molecule wave function are fixed by $r_{a1} = 0$ and 
$r_{b1} = a_{biex}$, (ii) one additional integral convolution with 
the ground-state excitonic envelope wave function is absent 
(the integration over $d {\bf r}_{a1}$). As a result, the 
authors end up with an incorrect conclusion, that ``the probability 
of the two-photon radiative transition is $\propto (a_{biex}/a_{ex})^6$, 
in contrast to the probability of the two-photon absorptive one 
which is $\propto (a_{biex}/a_{ex})^3$''. Furthermore, Eq.~(2) of 
Ref.~\cite{Gorban4} does not include the polariton effects, i.e.,  
the both crucial factors, the density of states and Hopfield's 
coefficients, are omitted. 

Thirdly, the expansion of the molecule wave function given by Eq.~(3) 
of Ref.~\cite{Gorban4} makes little sense, because (i) the biexciton 
wave function is assumed to be dependent only on two coordinates 
${\bf r}_{a1}$ and ${\bf r}_{b2}$, while the correct wave function 
should be specified by {\it three} independent vector coordinates, (ii) 
the r.h.s. of Eq.~(3) is not symmetric with respect to the permutation 
of two electrons or holes. The correct expansion of the molecule 
wave function is given by Eq.~(5) of Ref.~\cite{Tokunaga}. 

In order to assign the $M$-like emissions to the inverse exciton series 
one needs a severe critical analysis of the experimental data. While 
the observation of the $M$ series in CuCl requires a highly sensitive 
detection technique and a high-quality sample free from impurity 
emissions, our results are very stable and reproducible. Furthermore, 
${\bf k}$-selective excitation of excitonic molecules is crucial both 
for the final evidence of the observation of the $M$ series and for the 
reconstruction of the molecule wave function. The bipolariton model allows 
us to describe self-consistently a continuous, smooth change of the 
intensities and spectral positions of the $M_{ns}$ replicas with 
the total molecular wavevector ${\bf K}_m$ decreasing from $2 {\bf k}_0$ 
towards ${\bf K}_m = {\bf 0}$ (compare Fig.~1 and Fig.~3). 
From the criteria discussed above we conclude that Gorban {\it et al.} 
have not yet presented a convincing set of the experimental data 
on the $M$ series in $\beta$-ZnP$_2$.

\begin{figure}
\caption{ The photoluminescence spectra of a CuCl single crystal 
(sample $No.~1$) under the two-photon resonant generation of 
excitonic molecules ${\bf K}_m = 2 {\bf k}_0$ by the ps laser pump 
at 3.186~eV (solid curves) and under the interband excitation 
at 3.814~eV (dashed curves) with a cw HeCd Laser. }
\label{fig1}
\end{figure}

\begin{figure}
\caption{The inverse exciton series in the PL spectrum of a CuCl single 
crystal (sample $No.~2$) under the two-photon resonant excitation of 
molecules at ${\bf K}_m = 2 {\bf k}_0$.} 
\label{fig2}
\end{figure}

\begin{figure}
\caption{The inverse polariton series ($M$ series) from excitonic 
molecules resonantly created at ${\bf K}_m = {\bf 0}$.}
\label{fig3}
\end{figure}

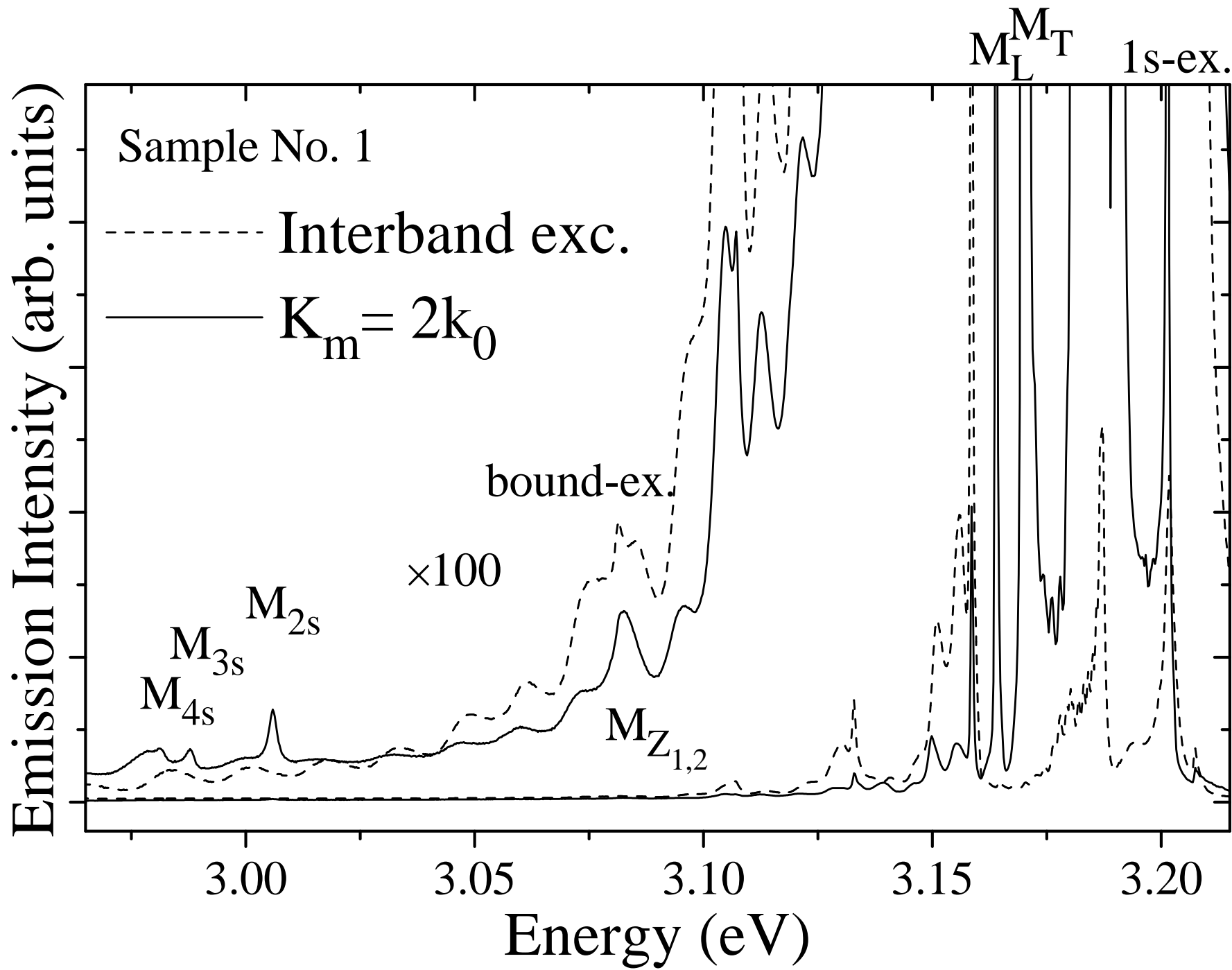
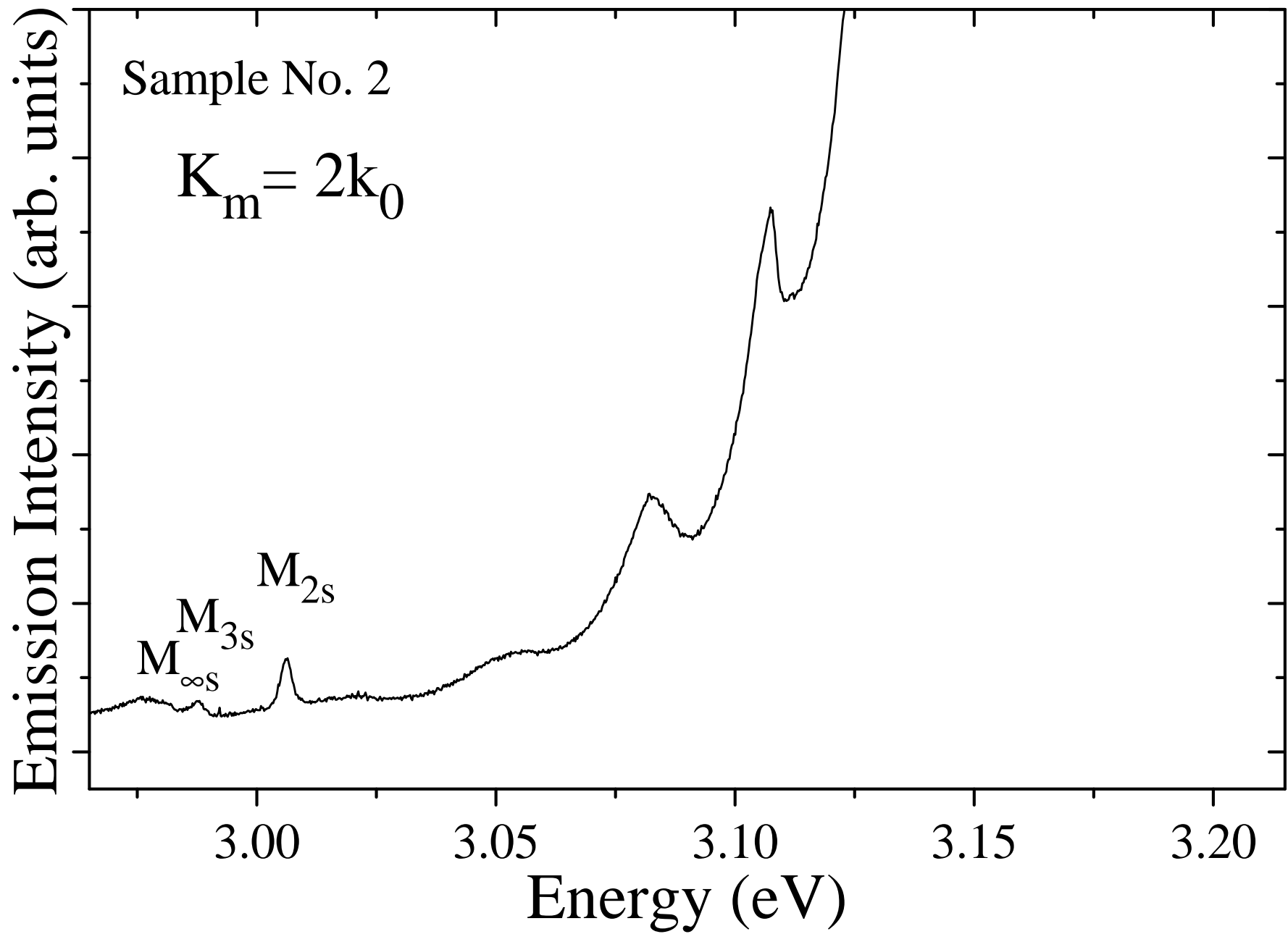
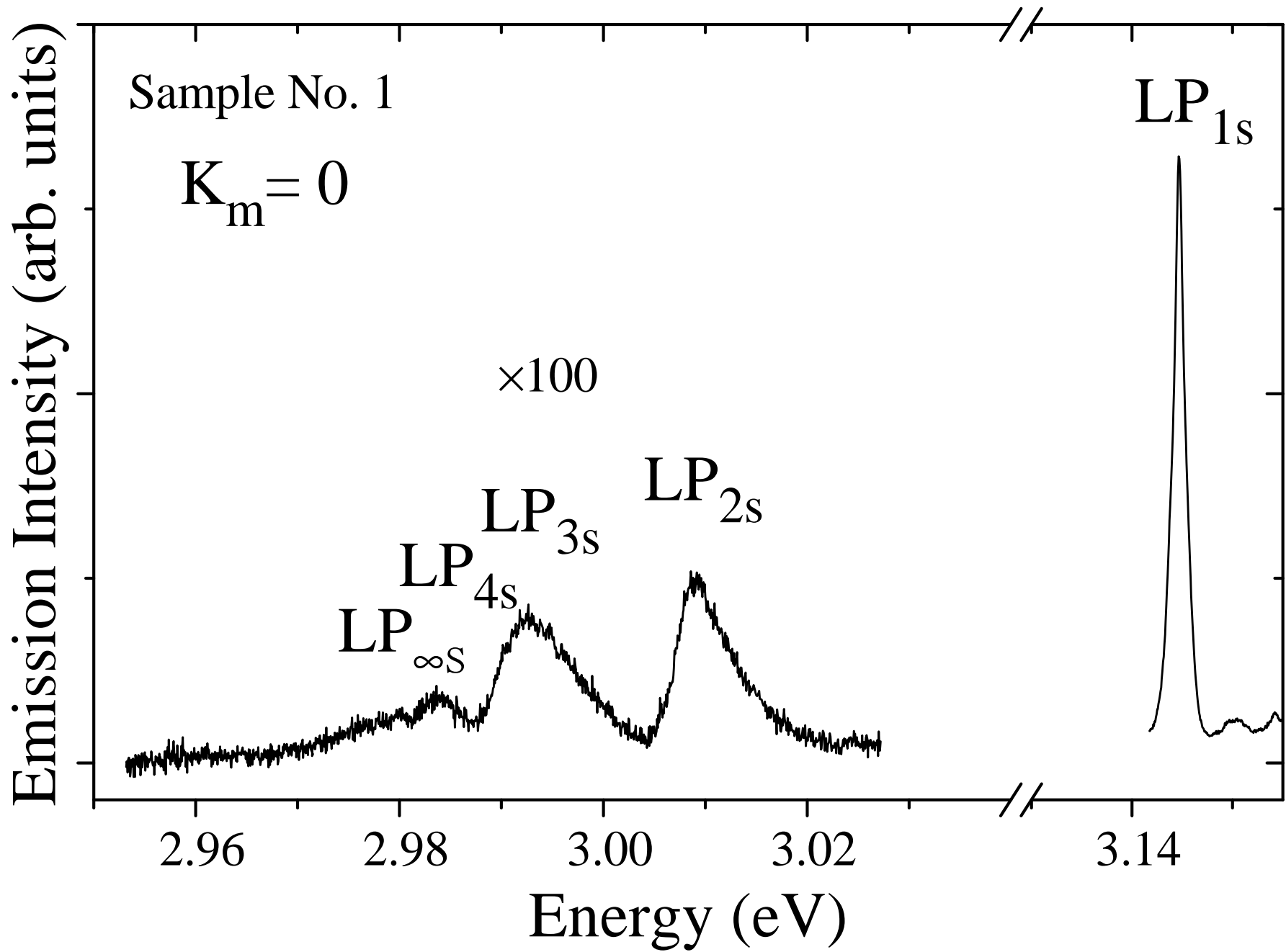
\end{document}